\documentclass[twocolumn,secnumarabic,amssymb, superscriptaddress, nobibnotes, aps, prl,showpacs]{revtex4-1}
\usepackage{graphics,graphicx,epstopdf,amsmath,natbib}

\pdfoutput=1

\begin{document}

\title{Resonant second-harmonic-generation circular-dichroism microscopy reveals molecular chirality in native biological tissues}%

\author{Mei-Yu Chen}%
\affiliation{Department of Physics, National Taiwan University, Taipei, Taiwan (R.O.C.)}
\author{Mikko J. Huttunen}%
\affiliation{Department of Physics and Max Planck Centre for Extreme and Quantum Photonics, University of Ottawa, Ottawa,  Ontario K1N 6N5, Canada}
\author{Che-Wei Kan}%
\affiliation{Department of Physics, National Taiwan University, Taipei, Taiwan (R.O.C.)}
\author{Yen-Yin Lin}%
\affiliation{Institute of Photonics Technologies, National Tsing Hua University, Hsinchu, Taiwan (R.O.C.)}
\affiliation{Department of Electrical Engineering, National Tsing Hua University, Hsinchu, Taiwan (R.O.C.)}
\author{Cin-Wei Ye}%
\affiliation{Department of Physics, National Taiwan Normal University, Taipei, Taiwan (R.O.C.)}
\author{Meng-Jer Wu}
\affiliation{Department of Physics, National Taiwan Normal University, Taipei, Taiwan (R.O.C.)}
\author{Hsiang-Lin Liu}%
\affiliation{Department of Physics, National Taiwan Normal University, Taipei, Taiwan (R.O.C.)}
\author{Shi-Wei Chu}%
\email{swchu@phys.ntu.edu.tw \\}
\affiliation{Department of Physics, National Taiwan University, Taipei, Taiwan (R.O.C.)}
\affiliation{Molecular Imaging Center, National Taiwan University, Taipei, Taiwan (R.O.C.)}

\begin{abstract}
Conventional linear optical activity effects are widely used for studying chiral materials. However, poor contrast and artifacts due to sample anisotropy limit the applicability of these methods. Here we demonstrate that nonlinear second-harmonic-generation circular dichroism spectral microscopy can overcome these limits. In intact collagenous tissues, clear spectral resonance is observed with sub-micrometer spatial resolution. By performing gradual protein denaturation studies, we show that the resonant responses are dominantly due to the molecular chirality. 
\end{abstract}


\maketitle

Many molecules exist in left- or right-handed forms that are mirror images of each other, similar to the human hands. This almost omnipresent property in nature is called chirality, and for example most biological molecules are chiral. Type I collagen, the most abundant protein in the human body, is no exception and has a special chiral molecular structure composed of three polypeptide strands in a left-handed conformation that twist together to form a right-handed triple helix configuration~\cite{shoulders2009}. Misfolding of the triple helices has been linked to various serious diseases~\cite{holmes1993,vogel1979a,willing1994,sykes1986osteogenesis}. Hence, it is important to be able to investigate the chiral properties of type I collagen. 

Conventionally, chiral materials are studied using linear optical activity (OA) effects, such as optical rotatory dispersion or circular dichroism (CD)\cite{barron2004}. 
Interestingly, the CD response can be enhanced near molecular resonances, providing the basis for utilizing CD spectroscopy as a powerful tool for studying secondary structures of proteins and to detect the denaturation of chiral polypeptides~\cite{charney1979}. The downside of CD spectroscopy is that  CD response requires light-matter interactions beyond the electric-dipole approximation, resulting in very weak relative signal strengths, i.e. contrast,  on the order of $10^{-3}$~\cite{Urbanova2007}. 
 
In  addition to linear OA effects, several nonlinear OA effects have been found, which can also provide information of the microscopic molecular structure~\cite{Fischer2005}.  Such effects include  vibrational circular dichroism~\cite{stephens1994ab}, Raman optical activity~\cite{barron2004}, two-photon-absorption circular dichroism~\cite{tinoco1975two,toro2010two}, sum-frequency generation circular dichroism~\cite{Fu2010} and second-harmonic-generation circular dichroism (SHG-CD)~\cite{petralli1993circular}. Among them, the highest so far reported contrast has been  reported for SHG-CD, where the relevant chiral information could be accessed by measuring the normalized differences of second-harmonic generation (SHG) between right-handed circularly (RHC) and left-handed circularly (LHC) polarized fundamental laser beam~\cite{sioncke2003second} 
\begin{equation} \label{Eq:SHG_CD}
\mathrm{SHG-CD} = \frac{I_{\mathrm{RHC}}(2\omega) -I_{\mathrm{LHC}}(2\omega)}{(I_{\mathrm{RHC}}(2\omega) +I_{\mathrm{LHC}}(2\omega))/2} \, .
\end{equation}
SHG-CD was first discovered using chiral binaphthol molecules adsorbed on an air–water interface~\cite{petralli1993circular} and was later demonstrated in molecular thin films~\cite{crawford1994second,byers1994electronic}. Previous studies also showed that SHG-CD can occur already within the electric-dipole approximation of the light-matter interaction explaining why its  relative strength  could be several orders of magnitude stronger than that of linear CD~\cite{petralli1993circular}. In addition, utilization of nonlinear processes, such as SHG-CD, results in intrinsic optical sectioning capabilities, as well as higher penetration depth, which are beneficial for potential three-dimensional chiral imaging applications~\cite{williams2005interpreting,kriech2005imaging,kissick2011second,lee2013chiral}.     

However, the origins of SHG-CD signals in intact biological tissues are not yet fully understood. Previous studies have shown that similar to linear OA effects, SHG-CD effects can also be influenced by possible sample anisotropy, challenging its use to study changes in protein secondary structures~\cite{hicks1994consequences,verbiest1996optical}. The issue of differentiating signals due to sample anisotropy and the actual molecular chirality can be especially challenging for real bio-tissues, since they are in practice also always anisotropic. Hsuan et al.~\cite{lee2013chiral} demonstrated that in a thick collagenous tissues, the contribution due to sample anisotropy (or orientation) of collagen molecules can dominate the SHG-CD responses for certain wavelengths. Nevertheless, an unambiguous imaging modality capable to differentiate between the possible contributions due to molecular chirality, e.g. protein conformation, and the macroscopic chirality due to sample anisotropy is a long-sought tool to study chiral systems with the capabilities of optical microscopy. Therefore, our central goal in this work is to study if SHG-CD spectroscopy could be used to solve the existing ambiguity, since the SHG-CD responses are, analogous to linear CD, expected to be enhanced at resonances related to the protein conformation~\cite{hicks1994consequences,verbiest1996optical,catone2012resonant,byers1994aa,byers1994ab,mesnil2002wavelength,xiao-Ou2010aa}. 

In this Letter, we demonstrate that in an intact biological tissue, SHG-CD spectroscopy can unambiguously provide information on the protein conformation when the excitation wavelength coincides with molecular resonances of the proteins. We perform spectral SHG-CD microscopy on collagenous tissues and show by gradually denaturing the sample, that although the non-resonant SHG-CD responses seem to be dominated by sample anisotropy, the resonant contribution to the SHG-CD is clearly due to molecular chirality i.e. related to protein conformation. We also validate our results by performing extensive numerical SHG-CD simulations showing very good agreement with the measurements. 

\begin{figure}[ht] 
\includegraphics[width=8.5cm]{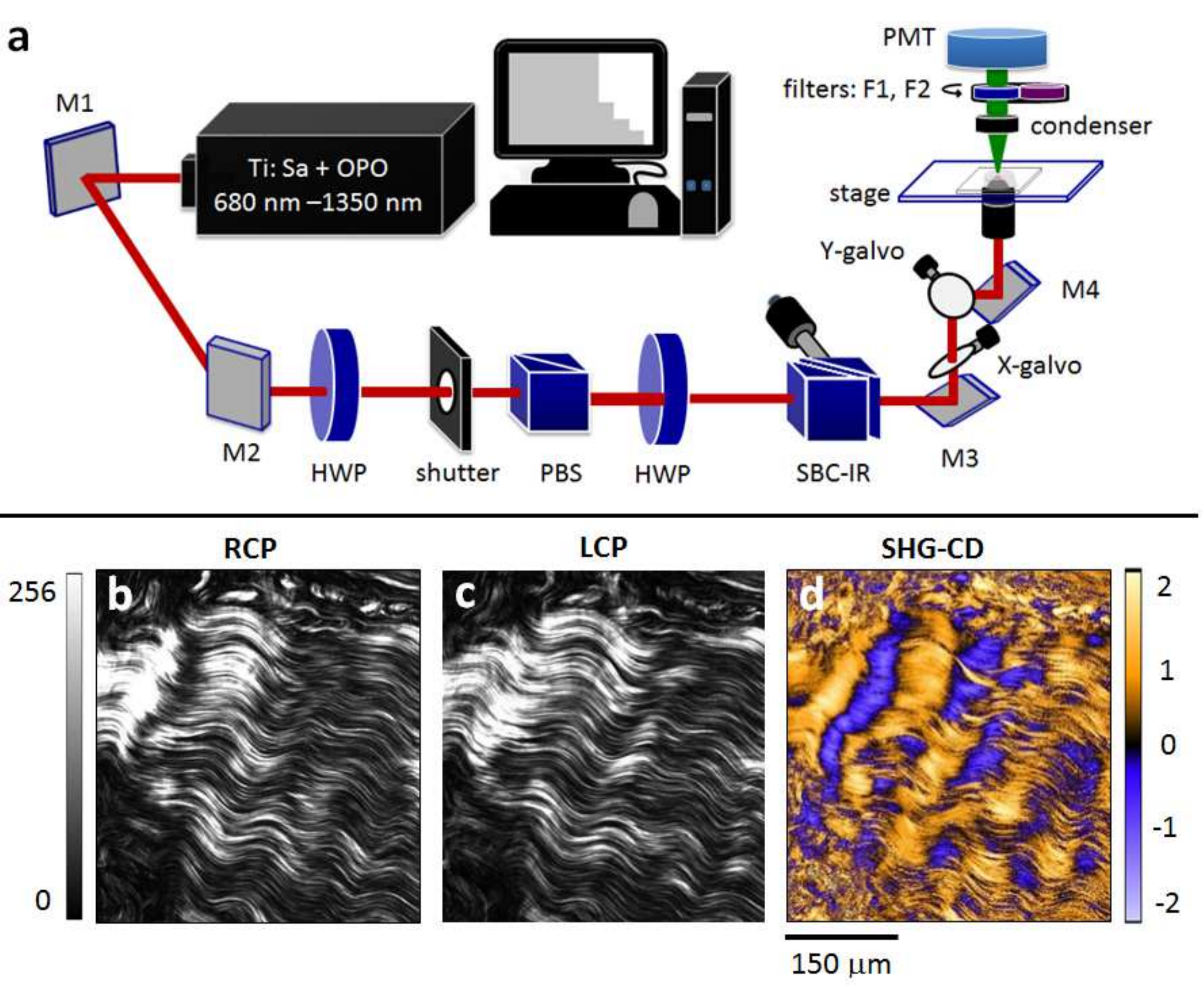} 
\caption{(a) Setup of a spectroscopic SHG-CD microscope. M1-M4: mirrors; HWP: half-wave plate; F1: short-pass filter with a 550~nm cutoff; F2: short-pass filter with an 890~nm cutoff. (b) and (c) are SHG-intensity images using RHC and LHC input polarization, respectively, at a wavelength of 1030 nm. (d) Corresponding SHG-CD image obtained from (b) and (c) by using Eq. (1).} 
\label{Fig:1}
\end{figure}

Our experimental setup is shown in Fig.~\ref{Fig:1}(a). A tunable Ti:sapphire femtosecond laser (Chameleon-Vision II, Coherent, USA) and an optical parametric oscillator (OPO) provided an excitation wavelength in the range of 680-1350~nm. A set of polarizing beam splitters (PBSs) and half-wave plates (HWPs) were used to maintain the power at 30 mW in the back aperture of the objective lens for the used wavelengths. The excitation beam was directed into a Leica TCS SP5 system with a pair of X-Y galvanometers to achieve raster scanning, and the beam was focused onto the specimen using an objective lens (HC PL IRAPO 20X, NA=0.75 water immersion, Leica). This specially designed objective lens has high transmission from visible to 1300~nm wavelength region and provides exceptional axial and lateral color-aberration correction throughout its spectral  range.  

The SHG signals were collected by a condenser in the forward direction and detected with a photomultiplier tube (PMT). A short-pass filter blocking wavelengths above 890~nm (FF01-890/SP-25, Semrock, USA) for 1100-1300~nm excitation and a one blocking wavelengths above 550~nm (SPF-550-0.50, CVI Laser Optics, USA) for 750-1050~nm excitation were placed in front of the PMT to ensure the effective suppression of excitation. A Soleil-Babinet compensator  (SBC-IR, Thorlabs, USA) was placed in front of the scanner to achieve high-quality circularly polarized light. For each wavelength, an optimal configuration of the SBC was determined to ensure that the ellipticity of circular polarization was less than 1.05 at the focus. The handedness of the input circular polarization was changed by rotating the half-wave plate in front of the SBC. This experimental setup allowed the study of the wavelength dependence of SHG-CD signals.  

We used pig tendons as our samples, since they are mainly composed of type I collagen known to give strong SHG-CD~\cite{lee2013chiral}. We used a Leica CM1950 clinical cryostat to prepare 20~$\mu$m-thick tendon sections, to show that the technique is not fundamentally limited to thin film or surface samples~\cite{kriech2005imaging,Fu2010}. The specimens were immersed in isotonic sodium-chloride solutions (each ml contained 9.0 mg of sodium chloride) and were each placed between a coverslip and a glass slide. The edges of the coverslips were sealed to minimize possible collagen denaturation due to  dehydration.  The linear  optical absorption of type I collagen was studied using a UV/VIS/NIR spectrometer (Lambda 900, PerkinElmer, USA). In the collagen denaturation study, the samples were placed on a custom-made graphite heater, where each sample was gradually heated from room temperature to  70$^{\circ}$C. The heating was performed in steps (40$^{\circ}$C, 50$^{\circ}$C, 60$^{\circ}$C, and 70$^{\circ}$C) where after achieving the target temperature a constant temperature was maintained   for 15 min before performing the SHG-CD measurements, followed by the next heating step. 

Figures ~\ref{Fig:1}(b) and ~\ref{Fig:1}(c) show SHG images from type I collagen using excitation wavelength of 1030~nm for RHC and LHC  polarized excitations, respectively. A characteristic crimping pattern of the collagen fibers is observed. It is obvious that the outlines of these two images are the same, but the actual intensity patterns are quite different. As defined by Eq.~(1),  SHG-CD can be calculated for each pixel from Figs.~\ref{Fig:1}(b) and ~\ref{Fig:1}(c),  resulting in SHG-CD image shown in Fig. 1(d). As expected, the SHG-CD signal reaches unity, which is much larger than that of conventional linear CD. However, both positive and negative SHG-CD values were observed. This does not imply that both right- and left-handed molecular chirality simultaneously exist in the collagen, since based on our previous study~\cite{lee2013chiral}, the positive and negative contrasts of SHG-CD are induced by the anisotropy and orientation of the collagen fibers, rather than molecular chirality. 

\begin{figure}[ht]         
\includegraphics[width=8cm]{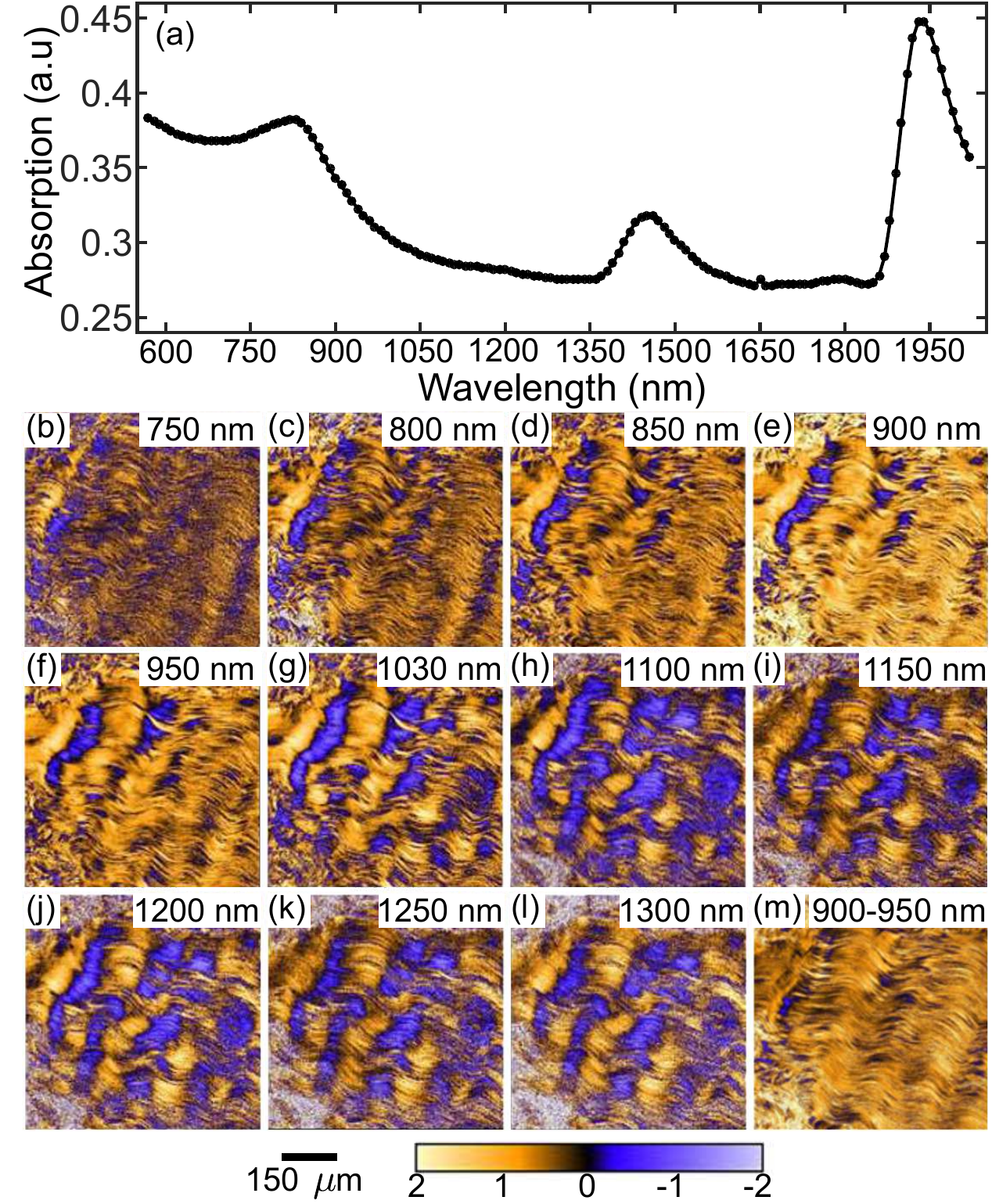} 
\caption{(a) Optical absorption spectrum of type I collagen tissue in the range of 600-2500 nm. (b-l) SHG-CD images for different excitation wavelengths in the range of 750-1300~nm. Positive, zero, and negative SHG-CD values are shown as orange, black, and blue, respectively. (m) The difference between the 900 and the 950~nm images show that the SHG-CD values are higher at resonance.}
\label{Fig:2}
\end{figure}

Figure ~\ref{Fig:2}(a) shows the optical absorption spectrum of type I collagen in the range of 600-2000~nm. In addition to the water absorption band around 1400~nm and 1900~nm~\cite{hale1973optical,Duke2012}, type I collagen exhibits an absorption band around 700-1000~nm~\cite{Taroni2009}, peaking near 875~nm. Because resonant behaviour of SHG-CD should be due to the molecular structure and not due to possible macroscopic sample anisotropy, we expected to see enhancement of  SHG-CD response near 875~nm resonance, while having a non-resonant background due to sample anisotropy. This hypothesis was also backed up by performing numerical SHG-CD simulations on sample structures resembling collagen fibers using an approach based on Ref.~\cite{Huttunen2015}, where the effects of molecular chirality and collagen crimping pattern on the expected SHG-CD were studied.   

\begin{figure}[ht]
\includegraphics[width=8cm]{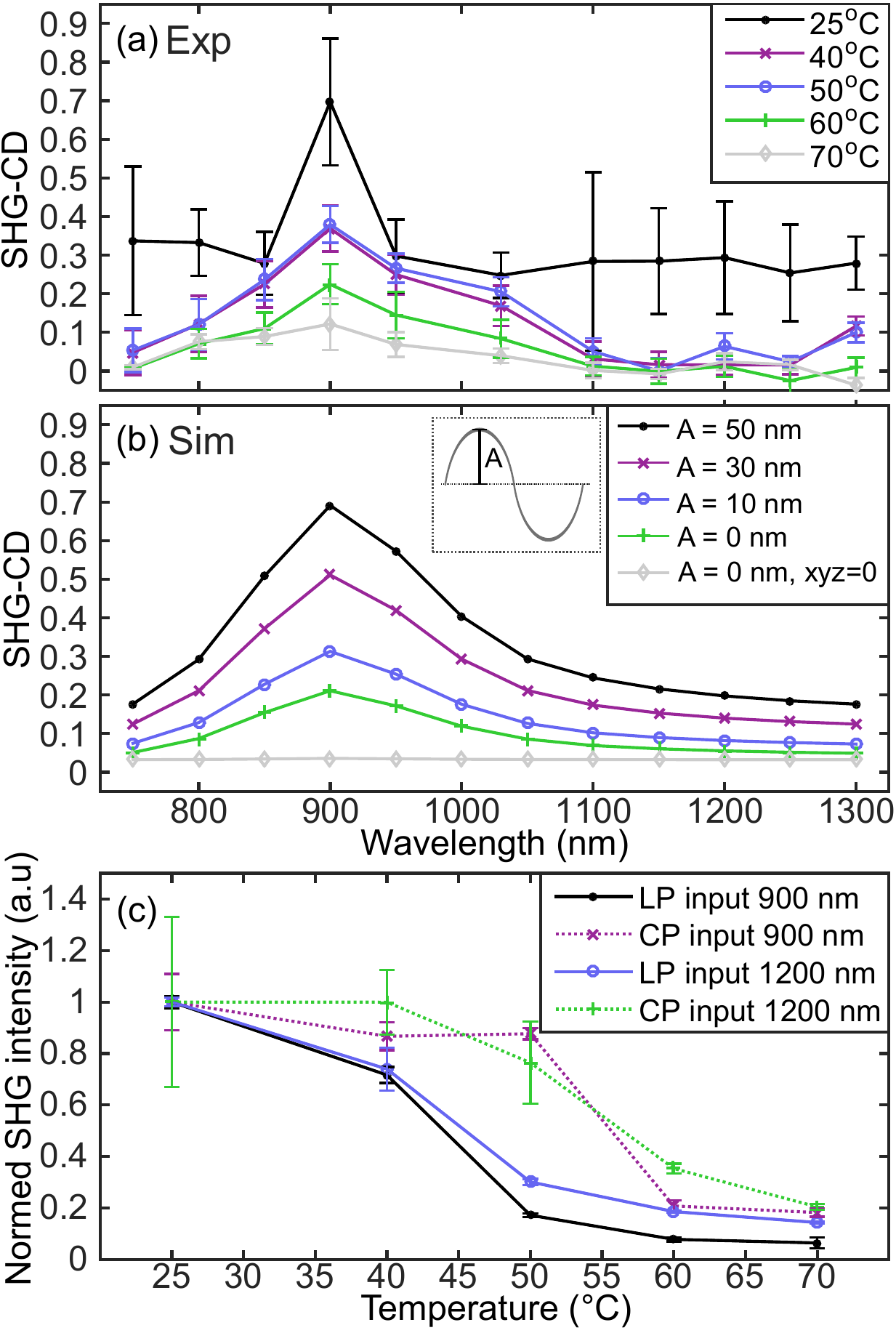} 
\caption{(a) Averaged SHG-CD spectra of type I collagen under gradual protein denaturation. First, the gradual heating removes the crimp pattern of collagen leading to decrease in the overall SHG-CD. The peak around 900~nm remains relatively unchanged, until the sample is heated over 53$^{\circ}$C, after which the secondary helical structure of collagen starts to denature. (b) Simulated SHG-CD responses while gradually decreasing the collagen crimping amplitude A from 50~nm to 0~nm. The role of molecular chirality was studied by decreasing the chiral susceptibility component $xyz$ to zero. (c) The overall SHG intensities also change when the sample is heated. Results are shown for fundamental wavelengths at 900 and 1200~nm and using linearly polarized (LP) and circularly polarized (CP) excitations. When the sample was heated to 50$^{\circ}$C, the overall SHG efficiency for LP input beam dropped to 25\% of the initial value while the overall SHG efficiency for CP input was still at ~90\%.}
\label{Fig:3}
\end{figure}

We measured a series of SHG-CD responses in the same focal plane with different excitation wavelengths (750-1300~nm), as shown in Fig.~\ref{Fig:2}(b-l). Clearly the SHG-CD image at 900~nm excitation appears brighter from the rest. In Fig.~\ref{Fig:3}(a), the black curve represents the wavelength dependence of the average SHG-CD of the entire image at room temperature (4 different samples). The average SHG-CD values dramatically increase at 900~nm, corresponding to the linear absorption peak of collagen in Fig. ~\ref{Fig:2}(a).  To our knowledge, this result is the first to unambiguously differentiate SHG-CD due to molecular chirality (protein conformation) from sample anisotropy. In addition, the measurements were performed on a complex microenvironment of intact biological tissue demonstrating the direct applicability of our results. 
		
The SHG-CD values are fairly stable $\sim$0.3 for non-resonant excitation wavelengths (700-800~nm, 1000 -1300~nm), as is seen on the black curve in Fig.~\ref{Fig:3}(a). We attribute this SHG-CD background to the average contribution from the anisotropy of collagen fibers. This anisotropy-induced SHG-CD corresponds well to the results of our previous work, where a 1040~nm femtosecond laser was used~\cite{lee2013chiral}. The bottom-right corner  of Fig.~\ref{Fig:3} shows the subtraction of the 900~nm and 950~nm SHG-CD images. The image color is almost completely orange, indicating that the SHG-CD values of most of the area increases when approaching resonance. This pixel-by-pixel analysis shows that although anisotropy-induced SHG-CD may be different at every microscopic pixel, positive SHG-CD contribution from molecular chirality dominates with resonant excitation. 

In order to further confirm our result, we  acquire SHG-CD spectra while gradually denaturing the proteins. The protein denaturation is performed by gradually heating the sample,  because heating is known to induce a conformational change in the collagen structure. It is known that collagen molecules dissociate at temperature range of 53-57$^{\circ}$C~\cite{Liao2011,Lin2005,Sun2006}. Therefore, we heated the collagen samples in steps from 25$^{\circ}$C to 70$^{\circ}$C, and measured a clear signal degradation in the SHG-CD spectra as shown in Fig.~\ref{Fig:3}(a). Closer inspection of Fig.~\ref{Fig:3}(a) shows, that the SHG-CD decreases in two parts  corresponding to different structural phases. In the first stage of heating, from 25$^{\circ}$C to 40$^{\circ}$C, the anisotropy-related SHG-CD value of approximately 0.3 disappears. This low-temperature heating affects mainly the macroscopic orientation and results in the vanishing of the crimp pattern of collagen fibers~\cite{Liao2011}. Nevertheless, the microscopic molecular triple-helix structure of the collagen proteins is almost unaffected at this temperature range~\cite{Sun2006}. Therefore, at the resonant wavelength of 900~nm, the molecular chirality contribution of SHG-CD ($\sim$0.4) is unraveled by this heating process at this temperature range. Then, in the range of 50$^{\circ}$C to 60$^{\circ}$C, the triple helix starts to denature, leading to diminished molecular chirality and significant decay of the SHG-CD resonant peak, while the SHG-CD values in the non-resonant regions remain approximately zero. Above 60$^{\circ}$C, further decay of the resonant peak is observed, reflecting the ongoing loss of the molecular chirality. 

The protein denaturation study agrees also well with the performed SHG-CD simulations. In brief, the used nonlinear parameters and the numerical approach follow closely Refs.~\cite{lee2013chiral,Huttunen2015}. The  molecular resonance peaked at 900~nm was assumed to follow a lorentzian lineshape with a life-time of 1~fs, which for simplicity was assumed to affect only the nonlinear susceptibility tensor components related to molecular chirality ($xyz=-yxz$). First we simulated spectral SHG-CD responses from an individual crimped collagen fiber for the used excitation wavelength range. Then repeated the simulations while gradually decreasing the crimp amplitude A from 50 to 0~nm. Finally, we studied the effect of protein denaturation on the SHG-CD by decreasing the strength of the susceptibility components related to molecular chirality. The simulation results are shown in Fig.~\ref{Fig:3}(b), and show a very good agreement with the experiments.

In addition to the SHG-CD measurements, it is well known that the SHG intensity of linearly polarized excitation is more sensitive to the anisotropy of collagen fibril orientation than  circularly polarized excitation~\cite{Chen2012}. Therefore, to further verify the existence of the anisotropy loss, the overall SHG intensities obtained with linearly polarized and circularly polarized excitations were measured as a function of the heating temperature, as shown in Fig.~\ref{Fig:3}(b). If the change during heating from 25$^{\circ}$C to 50$^{\circ}$C is primarily related to molecular anisotropy, SHG with linear polarization should drop faster than SHG with circular polarization. This is exactly what we observed in Fig.~\ref{Fig:3}(c). Interestingly, irrespective of whether the excitation wavelength is resonant (900~nm) or non-resonant (1200~nm), a similar tendency is observed for linear and circular polarizations, indicating that the anisotropy of the collagen molecules changes substantially during the low-temperature heating, and that the anisotropy effect is wavelength-independent. In addition, we note that although the anisotropy effect can be revealed via linearly polarized excitation, the chirality effect can only be probed via circularly polarized light. The conducted protein denaturation study provided clear evidence that the effects due to molecular chirality and more macroscopic chirality due to sample anisotropy on the overall  SHG-CD responses can be distinguished  by spectral SHG-CD measurements. 

In summary, we have performed spectral second-harmonic generation circular dichroism (SHG-CD) microcopy of type I collagen tissue with excitation wavelengths in the range of 750-1300 nm. We verified that two main mechanisms exist for SHG-CD in an intact biological tissue, macroscopic chirality due to sample anisotropy and molecular chirality due to protein conformation. We demonstrated that these two contributions can be differentiated when spectral SHG-CD responses are measured near resonant conditions, which we verified by observing different behavior for the resonant and non-resonant SHG-CD as a function of gradual protein denaturation by heating. Our results show that molecular chirality can be probed using resonant SHG-CD. For non-resonant wavelengths, SHG-CD imaging contrast provides information on the anisotropy. To our knowledge, this is the first successful attempt to probe molecular chirality from an original bio-tissue microenvironment. Therefore, our results opens new possibilities for studying chiral materials, such as biological tissues, at unprecedented spatial resolution using spectral SHG-CD microscopy. 

\subsection{Acknowledgments} This work was supported by the National Science Council of R.O.C. under Contract No. NSC-102-2112-M-002-018-MY3 and NSC-101-2923-M-002-001-MY3. SWC acknowledge the generous support from the Foundation for the Advancement of Outstanding Scholarship. 

\bibliography{Refs}

\end{document}